\documentclass[aps,secnumarabic,showpacs,nobibnotes,showkeys,
superscriptaddress,eqsecnum,amssymb,prd,amsfonts,nofootinbib,amsmath]{revtex4}
\usepackage{bm}
\begin{document}
\title{\Large\bf Quantising Gravity Using Physical States Of A Superstring}
\author{B. B. Deo}
\email{bdeo@iopb.res.in}
\affiliation{Department of Physics , Utkal University, 
Bhubaneswar-751004, India.}
\author{P. K. Jena}
\email{prasantajena@yahoo.com}
\affiliation{ Utkal University,Bhubaneswar-751004, India.}
\author{L. Maharana}
\email{lmaharan@iopb.res.in,lmaharan@yahoo.co.in}
\affiliation{Department of Physics , Utkal University, 
Bhubaneswar-751004, India.}
\begin{abstract}
A symmetric zero mass tensor of rank two is constructed using the 
superstring modes of excitation  which satisfies 
the physical state constraints of a superstring. These states 
have one to one correspondence with the quantised field operators and
are shown to be the absorption and emission quanta of 
the Minkowski space Lorentz tensor, 
using the quantum field theory method of quantisation. 
The principle of equivalence  
makes the tensor identical to the metric tensor 
at any arbitrary space-time point.
The propagator for the quantised field is deduced. 
The gravitational interaction 
is switched on by going over from ordinary 
derivatives to co-derivatives. The Riemann-Christoffel 
affine connections are calculated 
and the weak field Ricci tensor $R_{\mu \nu}^0$ 
is shown to vanish. The interaction part 
$R_{\mu \nu}^{int}$ is found out and the exact
$R_{\mu \nu}$ of the theory of gravity 
is expressed in terms of the quantised metric.
The quantum mechanical self energy of the 
gravitational field, in vacuum, is shown to
vanish. By the use of a projection operator, it is 
shown that the gravitons are the quanta of the general
relativity field which gives the Einstein equation $G_{\mu\nu}=0$.
It is suggested that quantum gravity may 
be renormalisable by the use of the massless
ground state of this superstring theory for general relativity
and a tachyonic vacuum creat and annihilate quanta of quantised 
gravitational field. 
\end{abstract}
\date{\today}
\maketitle
\section{Introduction}
Attempts to quantize and to have a renormalised Einstein's theory of 
gravitation have been a disappointing failure. The incompatibility of 
relativity with quantum mechanics was first pointed out by Heisenberg who 
commented that the usual renormalisation programme is ruined by the 
dimensional gravitational coupling constant. However, it is quite possible 
that the theory of gravity is finite to every order of its coupling. In 
order to achieve this, supergravity was pursued vigorously, but no success is 
yet in sight. On the other hand, soon after a theory of dual resonance 
model explaining all the postulates of S-matrix by a host of workers was put 
in place, Nambu\cite{Nambu70} and Goto~\cite{Goto71} worked out a classical 
relativistic string, which was raised to quantum level by Goldstone
\cite{Goldstein73}, Goddard, Rebbi and Thorn\cite{Thorn72} and 
also by Mandelstam\cite{Mandelstam75}. The very bold suggestion of Scherk and  
Schwarz \cite{Scherk74,Schwarz75} that the string theory carries  quantum 
information for all the four interactions including gravity, did not 
make much head way till 1984. Green and Schwarz \cite{Green84} formulated 
the superstring theory in ten dimensions which is still believed to be 
finite in all orders of perturbation theory. However, only later, it was found that
the heterotic string theory of  Gross, Harvey, Martinec and Rohm\cite{Gross85} 
is  the best candidate to explain gravitational interactions.

Casher, Englert, Nicolai and Taormina\cite{Casher85} have made a much publicised  
proof that the 26-dimensional bosonic string contains closed 10-dimensional 
superstrings, the two $ N=1$ heterotic strings and two $ N=2$ superstrings.  
This group have followed this up by making further incisive attempts to find 
a mechanism which generates space-time fermions out of bosons. To make contact 
with real physical world, one has to make the usual unsuccessful and nonunique 
compactification from ten to four dimensions. Kaku~\cite{Kaku98} and Green, 
Schwarz and Witten~\cite{Green87} in their books have rightly and clearly 
spelt out that `No one really knows how to break a 10-dimensional theory 
down to four'. In an earlier work~\cite{Bdeo06} on supergravity, using the 4-d 
superstring theory given below, one of the authors(BBD) deduced the propagator 
for the graviton. It was proved that the vanishing of the Ricci tensor using
the vielbeins in the tangent space formalism to go from flat to curved space 
time without using Riemann-Christoffel affine connections ( called affines 
in short) which are essential for a Quantum Theory of Gravity. Here we demonstrate 
this essential aspect. The affines come out in a simpler form and will 
simplify calculations of other quantum gravity problems.

In the earlier paper~\cite{Bdeo06}, one of us (BBD) also 
noted the important works of Feynman who used
gravity as a spin-2 field coupling to its energy 
momentum tensor. Mandelstam~\cite{Mandelstam75},Deser \cite{Deser76}
and DeWitt \cite{DeWitt67} have done extensive work in deriving
Feynman rules. The trouble is that gravity has too 
many constraints, inherent in the formal quantum gravity field 
theory. In the present approach, based on superstring,
most of the constraints have been taken care of in
constructing the superstring physical states. It is also 
important to realise that Riemann, Ricci, Weyl and all other
tensors of any relevance with quantum gravity are in
terms of general relativity metric tensor. This tensor is not
traceless. So the quantisation of the metric field strength
will have both the spin-2 graviton and spin-0 dilaton.
Eventhough, one of us(BBD) worked with pure graviton in
supergravity~\cite{Bdeo06}, we had to enlarge the scope, and approach 
the problem of quantum gravity by trying to quantise 
the metric tensor field. In all perturbative approaches, it is 
the general relativity metric tensor $g_{\mu\nu}(x)$ which is expanded
around the flat metric $\eta_{\mu\nu}$ as indicated in our concluding
section.

In our opinion, the best tool available to achieve a breakthrough is 
the superstring theory which, like gravity, needs to be formulated 
in the physical world of four dimensions. The simplest and the best 
way to descend directly from 26-dimensional  bosonic string to 
4-dimensional superstring is by using the Mandelstam equivalence 
between fermions and bosons in an anomaly free string theory.
The bosons are four in nature. The fermions belong to SO(3,1) bosonic
representation. They are divided into two groups. One group has 24 spinors
placed right handedly in six ways and the other 20 placed left handedly
in five ways.Thus the total number of the bosons is 4 and the fermions are 
4x6 and 4x5 have opposite handedness.
These will be relevant to gravity as much as the one of ref.~\cite{Casher85}. 
It is worth mentioning that our present construction of the superstring 
bears resemblance to the attempts by Gates et al\cite{Gates88}.

 The supersymmetric action, with $SO(6)\otimes SO(5)$ world sheet 
symmetry, turns out to be
\begin{equation}
S_{ss}=-\frac{1}{2\pi}~\int~d^2 \sigma 
\left [ \partial^{\alpha}X^{\mu}(\sigma,\tau)~
\partial_{\alpha}X_{\mu}(\sigma,\tau) - 
i\sum_{j=1}^6\bar{\psi}^{\mu,j} \rho^{\alpha}
\partial _{\alpha}\psi_{\mu,j} 
+ i\sum_{k=7}^{11}\bar{\phi}^{\mu,k} \rho^{\alpha}
\partial _{\alpha}\phi_{\mu,k}\right ],\label{eq1}
\end{equation}
with
\begin{eqnarray}
\partial_{\alpha}=(\partial_{\sigma},
~\partial_{\tau}),~~~
\rho^0 =
\left (
\begin{array}{cc}
0 & -i\\
i & 0\\
\end{array}
\right ),~~
\rho^1=
\left (
\begin{array}{cc}
0 & i\\
i& 0\\
\end{array}
\right )~~~\text{and}~~~
\bar{\phi}=\phi^{\dagger}\rho^0 .\label{eq2}
\end{eqnarray}
For the sake of completeness and to do justice to the subject of
gravity and strings, we briefly outline some details which have 
already been published elsewhere.

The  arrays ($e^j,~~e^k$) are the rows of ten zeros with only `1' in the
$j^{th}$ place or `$-$1' in the $k^{th}$ place. $e^je_j=6$ and $e^ke_k=5$. The 
invariation of the action is under the SUSY transformations with constraints 
to lead to spatial translations on two successive applications,
\begin{eqnarray}
\delta X^{\mu} &=& \bar{\epsilon} 
\left (e^j\psi^{\mu}_j - 
e^k\phi_k^{\mu}\right )=\bar{\epsilon} 
\Psi^{\mu},\label{eq3}\\
\delta\psi^{\mu,j}&=& -i\epsilon~ e^j\rho^{\alpha}
\partial_{\alpha} X^{\mu},~~~
\psi^{\mu}_j=e_j\Psi^{\mu},\label{eq4}
\end{eqnarray}
and
\begin{equation}
\delta\phi^{\mu,k}= -i\epsilon~ 
e^k\rho^{\alpha}\partial_{\alpha} X^{\mu},~~~
\phi^{\mu}_k=e_k\Psi^{\mu}.\label{eq5}
\end{equation}
Here $\epsilon$ is a constant anticommuting spinor and 
\begin{equation}
\Psi^{\mu}=e^j\psi^{\mu}_j - e^k\phi^{\mu}_k, \label{eq6}
\end{equation}
is the superpartner of $X^{\mu}$. This emits 
quanta of $\psi^{\mu}_j$ or $\phi^{\mu}_k$
while in the site $j$ or $k$ respectively. The string fields 
are quantised for the coordinates
\begin{equation}
X^{\mu}(\sigma,\tau)= x^{\mu} +p^{\mu}\tau + 
i\sum_{n\neq0}\frac{1}{n} \alpha_n^{\mu}~
e^{-in\tau}~\cos(n\sigma). \label{eq7}
\end{equation}
In terms of complex coordinates $z=\sigma + i\tau$ and 
$\bar{z}=\sigma - i\tau$, we have,
\begin{equation}
X^{\mu}(z,\bar{z})= x^{\mu} -i\alpha^{\mu}_0\ln |z| 
+i\sum_{m\neq 0}\frac{1}{m} \alpha_m^{\mu}~z^{-m}. \label{eq7a}
\end{equation}
Further,
\begin{equation}
\psi_{\pm}^{\mu,j}(\sigma,\tau)=
\frac{1}{\sqrt{2}}\sum_{r\in Z+\frac{1}{2}}
b^{\mu,j}_r~e^{-ir(\sigma \pm\tau)},~~~~~
 ~~~~~\phi_{\pm}^{\mu,k}(\sigma,\tau)=
\frac{1}{\sqrt{2}}\sum_{r\in Z+\frac{1}{2}}
b^{'\mu,k}_r~e^{-ir(\sigma \pm\tau)}
~~~ \text{for ~~~ NS~~ sector},\label{eq8}
\end{equation}
and 
\begin{equation}
\psi_{\pm}^{\mu,j}(\sigma,\tau)=
\frac{1}{\sqrt{2}}\sum_{m=-\infty}^{\infty}
d^{\mu,j}_m~e^{-im(\sigma \pm\tau)},~~~~
~~~~~\phi_{\pm}^{\mu,k}(\sigma,\tau)
=\frac{1}{\sqrt{2}}\sum_{m=-\infty}^{\infty}
d^{'\mu,k}_m~e^{-im(\sigma \pm\tau)}~~ 
\text{for~~ R~~sector}.\label{eq11}
\end{equation}
The bosonic quanta obey the commutation relation and the fermionic quanta obey the
anticommutation relation with the only major difference for Majorana 
fermions $b_{-r}=b^{\dag}_r$ but $b'_{-r}=-b^{'\dag}_r$ and $d_{-r}=d^{\dag}_r$  but 
$d'_{-r}=-d^{'\dag}_r$. So the number level density $N_B$ (for bosons) and $N_F$ (for
fermions) are
\begin{equation}
N_B=\sum_{\mu}\langle \phi|\alpha^{\mu}_{-1}\alpha_{1\mu}|\phi \rangle=4,
\end{equation}
and
\begin{equation}
N_F=\sum_{\mu}\sum_{i=1}^6 \langle \phi|b^{\mu}_{-i}b_{i\mu}|\phi \rangle
+\sum_{\mu}\sum_{j=1}^5 \langle \phi|b^{'\mu}_{-i}b^{'}_{i\mu}|\phi \rangle 
=\sum_{\mu}\sum_{i=1}^6 \langle \phi|b^{\dag \mu}_{i}b_{i\mu}|\phi \rangle
-\sum_{\mu}\sum_{j=1}^5 \langle \phi|b^{'\dag \mu}_{i}b^{'}_{i\mu}|\phi \rangle 
=24-20=4.
\end{equation}
These number level densities are as required by supersymmetry.

It may be apprehended that the assembly of forty four fermions may give rise to
a spectrum in space time,  will be highly pathological. However, of the 44, 
when 24 Majorana fermions are excited in one way, the other 20 are excited 
in the opposite way.Therefore only four Majorana fermions are effective.
It will show less complexity in pathology than an assembly of ten fermions in 
10-D superstrings.

 In the light cone basis, the energy momentum tensors $T_{++}$, $T_{--}$ and 
the currents $J_+ , J_-$ are   given by
\begin{eqnarray}
T_{++}&=&\partial_+X^{\mu}\partial_+X_{\mu} 
+\frac{i}{2}\bar{\psi}_+^{\mu,j}
\partial_+\psi_{+\mu,j} -\frac{i}{2}\bar{\phi}_+^{\mu,k}
\partial_+\phi_{+\mu,k},\label{eq11a}\\
T_{--}&=&\partial_-X^{\mu}\partial_-X_{\mu} 
+\frac{i}{2}\bar{\psi}_-^{\mu,j}
\partial_-\psi_{-\mu,j} -\frac{i}{2}\bar{\phi}_-^{\mu,k}
\partial_-\phi_{-\mu,k},\label{eq12}\\
J_+&=&\partial_+X_{\mu}\Psi_+^{\mu},\label{eq12a}\\
\text{and}~~~~~   J_-&=&\partial_-X_{\mu}\Psi_-^{\mu}\label{eq13}
\end{eqnarray}
 There are also energy momentum tensors associated 
with conformal $T_{++}^{FP}$
and superconformal ghosts $T_{++}^{SC}$ with 
their generators. Their corresponding quanta are
\begin{eqnarray}
T_{++}^{FP}&=&\frac{1}{2}c^+\partial_+b_{++} 
+\partial_+c^+b_{++},~~~~~~L_m^{FP}=
\sum_m^{FP}(m-n)b_{m+n}c_{-n}\\
T_{++}^{SC}&=&-\frac{1}{4}\gamma\partial_+\beta 
-\frac{3}{4}\beta\partial_+\gamma,~~~~~~\text{and}~~~~~
L_m^{gh,sc}=\sum_m^{SC}(\frac{1}{2}m+n)
:\beta_{m-n}\gamma_{n}:.\label{eq14}
\end{eqnarray}
Here $(b,c)$ obey the anticommutation rules and $(\beta, \gamma)$ 
obey the commutation rules. 

The superconformal ghost action follows from the local 
fermionic symmetry of the superconformal invariance 
of the action as used by Brink, De Vecchia, Howe, 
Deser and Zumino~\cite{brink76},
\begin{equation}
\delta \chi_{\alpha}=i\rho_{\alpha}\eta,
~~\text{and}~~~\delta e^a_{\alpha}=
\delta\Psi^{\mu}=\delta X^{\mu}=0,
\end{equation}
where $\eta$ is an arbitrary Majorana spinor and $e^a_{\alpha}$
is the usual `Zweibein', so that the gravitino can be gauged away using
\[
\delta \chi_{\alpha}=i\rho_{\alpha}\eta + \nabla_{\alpha}\epsilon .\]
Using the variation of $\chi, X_{\pm}$ and following 
reference~\cite{brink76}, one finds (\ref{eq14}). The
central charge `c' is given by c~=~1$-3k^2$. 
The  anomaly free total super 
Virasoro generator  can be written in two 
equivalent ways, each having zero central charge
\begin{equation}
L^{total}_m=L_m^{FP} + \frac{1}{\pi}
\int_{-\pi}^{\pi} e^{im\sigma}T_{++} d\sigma
=L_m^{FP}+L_m^{SC}+\frac{1}{\pi}
\int_{-\pi}^{\pi} e^{im\sigma}(T_{++}-k\partial_+
T^F_+) d\sigma,\label{eq15}
\end{equation}
where 
\begin{equation}
T_+^F=\frac{i}{2}\left ( \psi^{\mu,j}_+\psi_{+\mu,j}-
\phi^{\mu,j}_+\phi_{+\mu,j} \right ).
\label{eq16}
\end{equation}
The quantity  $k$ is related to the 
conformal dimension $J=\frac{1+k}{2}$
and equals $\frac{1}{\sqrt{6}}$ here. The normal 
ordering constant $\it{a}=-$1 in either case.

The super Virasoro generators $L_m$ of energy momenta  and the currents 
$(G,F)$ are~\cite{Bdeo06}

\begin{eqnarray}
L_m&=&\frac{1}{2}\sum^{\infty}_{-\infty}
:\alpha_{-n}\cdot\alpha_{m+n}:
+\frac{1}{2}\sum_{r\in z+\frac{1}{2}}(r+\frac{1}{2}m): 
(b_{-r} \cdot b_{m+r} - b_{-r}' \cdot b_{m+r}'):~~~~\text{NS},\label{eq16a}\\
L_m&=&\frac{1}{2}\sum^{\infty}_{-\infty}
:\alpha_{-n}\cdot\alpha_{m+n}:
+\frac{1}{2}\sum^{\infty}_{n=
-\infty}(n+\frac{1}{2}m): (d_{-n} \cdot d_{m+n} - d_{-n}'
\cdot d_{m+n}'):~~~~~\text{R},\label{eq17}\\
G_r &=&\frac{\sqrt{2}}{\pi}\int_{-\pi}^{\pi}
d\sigma e^{ir\sigma}J_{+}=
\sum_{n=-\infty}^{\infty}\alpha_{-n}\cdot 
(e^j b_{r+n,j}^{\mu} - e^k b_{r+n,k}^{\prime\mu}),\\
\text{and}~~~~~F_m &=&\sum_{n=-\infty}^{\infty} 
\alpha_{-n}\cdot (e^j d_{m+n,j}^{\mu} 
- e^k d_{m+n,k}^{\prime\mu}).\label{eq18}
\end{eqnarray}

These satisfy the super Virasoro algebra with 
central charge $c=26$. The second version of equation (\ref{eq15})
can be easily written down\cite{Kaku98,Green87}.

The physical states are defined through the relations
\begin{eqnarray}
(L_0-1)|\phi\rangle&=&0,~~~L_m|\phi\rangle=0,~~~G_r|\phi\rangle=0,~~  
~~~~~~~~~~\text{for}~~(r,m)~ >~ 0,~ \text{NS ~~(bosonic)}\label{eq19}\\
(L_0-1)|\psi\rangle_{\alpha}&=&(F^2_0-1)|\psi\rangle_{\alpha},
~~~L_m|\psi\rangle_{\alpha}=F_m|\psi\rangle_{\alpha}=0,
,~~~\text{for}~~~m ~ >~0,~~\text{R~~ (fermionic)}\label{eq20}
\end{eqnarray}
and the mass spectrum is given as
\begin{eqnarray}
\alpha'M^2&=& -1,~-\frac{1}{2},~ 0 ,~ \frac{1}{2},~ 1,~
\frac{3}{2},...~~~~~~~~~~~~~~\text{NS}\\
\text{and}~~~~~~~~~~~~~~~~~~~~~~~~~~~&&\nonumber \\
\alpha'M^2&=& -1,~ 0 ,~ 1,~ 2,~ 3,...~~~~~~~~~~~~~~~~~\text{R}.
\end{eqnarray}
The G.S.O. project out~\cite{Gliozi77} the half integral mass spectrum 
values, tachyonic or otherwise. The tachyonic bosonic energy of the NS
sector  $\langle0| (L_0-1)^{-1}|0\rangle_S$ is cancelled by   
$\langle0| (F_0-1)^{-1}(F_0+1)^{-1}|0\rangle_R$ of the Ramond sector. 
This is typical of supersymmetry and is well known. Just to remind, 
the SUSY charge is~\cite{Deo05},
\begin{equation}
Q=\frac{1}{\pi}\int d\sigma\rho^o \rho^{+\alpha}\partial_{\alpha}
X^{\mu}\Psi_{\mu},\label{eq20a}
\end{equation}
and since, the Hamiltonian is related as
\begin{equation}
\sum_{\alpha}\{Q_{\alpha}^{\dag},~Q_{\alpha}\}=2H,
~~~~\text{and}~~~~ \sum_{\alpha}|Q_{\alpha}|^2=2\langle
\phi_0|H|\phi_0\rangle,\label{eq20b}
\end{equation}
the ground state is massless and the physical  Fock space is tachyonless.
The references~\cite{Deo05,Chattaraputi02} give a detailed derivation of 
the modular invariance and vanishing the partition function of the string.
Thus with no anomaly in four dimensional superstring theory, we can go over
to construct zero mass Fock space physical state functions. We shall 
demonstrate this in the next sections and attempt to quantise gravity using 
ideas based on superstring theory and its exact equivalence with quantum
operators.  

\section{String states $\longleftrightarrow$ quantum operators}

In the bosonic superstring, the Neveu Schwarz bosonic sector contains
the bosonic tachyon. This tachyon state is very useful to construct
the ground state of zero mass. The vacuum state $|0,0 \rangle$ of the string
is the functional integral of the string theory over a semi-infinite strip.
This could be conformally mapped to the unit circle. These ideas become clearer,
if we consider a closed string where we have a second set of Virasoro
algebraic equations. The following is a recipe, given by Polchinski
~\cite{Polchinski98}, prescribing the link between superstring states and operators,
which is very important for quantising gravity.

Radial quantisation has a natural isomorphism between the string state space
of conformal field theory (CFT) in a periodic spatial dimension, and the space 
of local operators~\cite{Polchinski94}. Let there be a local isolated operator
$\cal{A}$ at the origin and no more inside the unit circle denoted by $|z|=1$, but
with no other specification outside the circle. Let us open a slit in the circle
and consider the path integral on the unit circle, giving an inner product
$\langle \psi_{out}|\psi_{in} \rangle$. Here, $\psi_{in}$ is the incoming state 
given by the path integral $|z|<1$ and $\psi_{out}$ is the outgoing one at
$|z|>1$. Explicitly, a field $\phi$ is split into integrals outside, inside and
on the circle. The last one will be called $\phi_B$. The outside integral is
${\psi_{out}(\phi_B)}$ and the inside integral is ${\psi_{in}(\phi_B)}$, and
the remainder is $\int[d\phi_B]~\psi_{out}(\phi_B)\psi_{in}(\phi_B)$. 
The incoming state depends on the operator $\cal{A}$ and hence is denoted by
$|\psi_A \rangle$. This is the needed important mapping from operators to 
states. Summarising, `the mapping from operators to states is given by the path
integral on the unit disk'. The inverse is also true.

If $Q$ is any conserved charge, $Q|\psi_A \rangle$ is the operator equivalent 
of $Q \cdot \cal{A}$. In particular, if $\cal{A}$ is the unit operator $\openone$, 
and 
$Q=\alpha_m=\oint \frac{dz}{2\pi}z^m \partial X$, for $m \geq 0$, so that 
$\partial X$ is analytic and the integral vanishes for $m \geq 0$, we get 
$\alpha_m|\psi_{\openone}\rangle=0$,~~ $m \geq 0$. This establishes the 
exact correspondence of the unit operator to the string vacuum,
\begin{equation}
\openone \leftrightarrow |0,0 \rangle.
\end{equation}
Similarly, one finds the operator equivalence,
\begin{equation}
:e^{ik.X(z)}:~ \leftrightarrow |0,k \rangle.\label{eq2a}
\end{equation}
In the above, $X(z)$ is given by equation (\ref{eq7a}). $:e^{ikX}:$ implies 
normal ordering of the operators contained in it.
The first number in the bra refers to $m$, and second to the eigenvalue
of $\alpha_o^{\mu}$ i.e. $\alpha_o^{\mu}|0,k\rangle=k^{\mu}|o,k\rangle$. So for
the tachyon, $|0,k\rangle\leftrightarrow e^{ik.x}$, because on the 
circumference of the circle $|z|$=1, the tachyonic vacuum can not
annihilate\cite{Kaku98}. This equivalence is utilised to convert the string 
states tensor of rank two metric field of graviton and dilaton to quantum 
operators and vice versa. The CFT unitarity gives

\begin{equation} 
\langle 0,k|0,k^{\prime} \rangle = 2 \pi~ \delta (k-k^{\prime}).
\end{equation}
The three spatial components would lead to
\[
\langle 0,\vec k|0,\vec k^{\prime} \rangle = (2 \pi)^3~ \delta^{(3)}
 (k-k^{\prime}).\label{eq2b}
\]

This is generalised to normalisation of massless states with $k_0=|\vec k|$ and 
we use one like the massive vector meson renormalisation,
\begin{equation}
\langle 0,\vec k|0,\vec k^{\prime} \rangle = (2 \pi)^3~(2k_0)~ \delta^{(3)}
 (k-k^{\prime}).\label{eq2c}
\end{equation}

\section{ Quantisation of  the gravitational field metric using superstring states}

To construct the second rank Lorentz tensor of general relativity, we shall use 
the quanta $b^{\mu}_i$ of the NS bosonic sector which comes from the SO(3,1) 
$\psi^{\mu}_j$'s, belonging to the SO(6) group of the action (\ref{eq1}). 
In the NS sector, the hidden tachyonic vacuum $|\phi\rangle=|0,k\rangle$
of momentum $k$ is such that $(L_0-1)|0,k\rangle=0$. This satisfies the 
superstring Virasoro physical constraint of equation (\ref{eq13}). The ghost free
physical Fock space states containing matter and radiation are built up by
operating creation operators on this state. For quantum gravity, we need massless
quanta of spin 2 and a metric tensor. The later when quantised, would have
both massless quanta of spin 2 and spin 0. Departing from earlier approaches 
to quantum gravity, we first proceed to construct a
metric tensor quantum operator. This should be a Lorentz tensor of rank 2.
The Lorentz tensor, which is symmetric but not traceless, is simply given by
the quantum operator

\begin{equation}
g_{\mu\nu}(k)= ~:\sum_{ij}c_{ij}b^{i\dag}_{\mu} b^{j\dag}_{\nu}  
e^{ik.X}:~=\sum_{ij}c_{ij}b^{i\dag}_{\mu}b^{j\dag}_{\nu}
:e^{ik.X}:~\longleftrightarrow a^{\mu \nu}|0,k\rangle=
a^{\mu\nu}~e^{ik.x} ;~~ c_{ij}=-c_{ji},\label{eq21}
\end{equation}
where the operator $a_{\mu \nu}$ is,

\begin{equation}
a_{\mu \nu}=\sum_{ij}c_{ij}b^{i\dag}_{\mu}b^{j\dag}_{\nu}, \label{eq21a}
\end{equation}
and commutes with $\alpha_{\mu}$'s,  we have used the operator $\longleftrightarrow$ 
state equivalence of equation (\ref{eq2a}). This operator creates a pair of 
fermions. Since 

\begin{equation}
[L_0, b^{\dagger\nu}_{j}]=\frac{1}{2}
b^{\dagger\nu}_{j}\label{eq22}
\end{equation}
and
\begin{equation}
[G_{\frac{1}{2}}, b^{j\nu\dag}]= \alpha_{0}^{\nu}~e^j=
k^{\nu}e^j,\label{eq23}
\end{equation}
we have,
\begin{equation}
L_0~g_{\mu\nu}(k)= L_0\sum_{ij}c_{ij}
b^{i\dag}_{\mu}b^{j\dag}_{\nu} :e^{ik.X}: \longleftrightarrow \sum_{ij}c_{ij}
b^{i\dag}_{\mu}b^{j\dag}_{\nu} (L_0-1)|0,k\rangle=0,\label{eq24}
\end{equation}
since $(L_0-1)|0,k\rangle=0$ as stated. So the quantum operator $g_{\mu\nu}(k)$ 
is the ground state which is a  pair of vectorial quanta created, would lead to 
a massless state. In order to satisfy the G.S.O. condition, we examine  
equation (\ref{eq21}) 
\begin{eqnarray}
\left (1+(-1)^F\right )G_{\frac{1}{2}}g_{\mu\nu}(k)&=&\left (1+(-1)^F\right )
\left [G_{\frac{1}{2}}, g_{\mu\nu}(k)
\right ]\label{eq25}\\
&=&\left (1+(-1)^F\right )\frac{1}{\sqrt 2} \sum c_{ij}\left (e^ik_{\mu} 
b_{\nu}^j - e^jk_{\nu} b_{\mu}^{i}\right )|0,k\rangle=0
\label{eq26}\\
&=&k_{\mu}\sum c_{ij}~e^i b^{j\dag}_{\nu}|0,k\rangle.\label{eq27}
\end{eqnarray}
Since $G_{\frac{1}{2}}$ annihilates one fermion of the pair, the 
lone left out fermion is G.S.O. projected out.
So $g_{\mu\nu}(k)$ is a ground state with zero energy.

Operator $a_{\mu \nu}$ defined in equation(\ref{eq21a}) is seen to satisfy 
the relations,
\begin{equation}
\left [  a_{\mu\nu}, ~a_{\lambda\sigma}^{\dag}\right ]
=f_{\mu\nu,\lambda\sigma}|c|^2,~~\left [  a_{\mu\nu}, ~a_{\lambda\sigma}\right ]=
\left [  a_{\mu\nu}^{\dag}, ~a_{\lambda\sigma}^{\dag}\right ]=0, \label{eq29}
\end{equation}
where  
\begin{equation}
f_{\mu\nu,\lambda\sigma}=\eta_{\mu\lambda}\eta_{\nu\sigma}
+\eta_{\mu\sigma}\eta_{\nu\lambda}.
\end{equation}
Here the flat space metric $\eta_{\mu \nu} $ is, as usual,
\begin{eqnarray}
\eta_{\mu \nu}=
\left (
\begin{array}{cccc}
-1 & 0 & 0 & 0\\
0 & 1 & 0 & 0\\
0 & 0 & 1 & 0\\
0 & 0 & 0 & 1
\end{array}
\right ).
\end{eqnarray}
For the desired normalisation for $|c|^2$, in equation(\ref{eq29}) will be 
suitably chosen.

In the Gupta-Bleuler method of quantisation~\cite{Gupta50} of
only the ground state symmetric tensor
$g_{\mu \nu}(k)$. The state satisfies the equation
\begin{equation}
L_0~|~g_{\mu\nu}(k)~\rangle = L_0~a_{\mu\nu}^{\dag}:e^{ikX}:~\leftrightarrow 
L_0~a^{\dag}_{\mu\nu}|0,k\rangle=0\leftrightarrow L_0~a^{\dag}_{\mu\nu}~e^{ik.x}.\label{g1}
\end{equation}
Likewise \[ :e^{-ikX}a_{\mu\nu}^{\dag}:~~~ \longleftrightarrow ~~\langle 0,k|a_{\mu\nu}. \]
Consider the expansion of $ {g}_{\mu \nu}(x)$, given by
\begin{equation}
{g}_{\mu \nu}(x)= \int \frac{d^4 k}{(2 \pi)^4} 
\left ( a^{\dag}_{\mu \nu}~e^{ikx} + a_{\mu \nu}~e^{-ikx}\right ).\label{g2}
\end{equation}
which satisfies the zero mass Klein-Gordon equation
$ \Box {g}_{\mu \nu}(x)=0.$ The  collection of 
states (\ref{g1},~\ref{g2}), due to the isomorphism with operators,
is the field quantisable as, for instance, in Gupta-Bleuler formalism~\cite{Gupta50}.
When ${g}_{\mu \nu}(x)$ is eventually quantised,
the associated creation and annihilation operators correspond to
massless quanta. It may be pointed out that this is same as the 
Gupta-Bleuler formalism which is covariant 
and hence used in most string quantisations to maintain manifest 
Lorentz covariance. Thus the string Fock states $ g_{\mu\nu}(k)$ and 
the hermitian conjugates will turn out to be related to the quantum creation 
and annihilation quanta of spin-2 and spin-0  for the field tensor.
In the instance case, $a^{\dag}_{\mu \nu}(k)$ and $a_{\mu \nu}(k)$ are the 
creation and annhilation quanta of these objects.

Before proceeding further, we must ensure that $g_{\mu\nu}(k)$ 
is a special type of symmetric Lorentz tensor \cite{Weinberg72} which, 
under a proper Lorentz transformation from $x$ to $x^{\prime}$, 
does not behave as the usual second rank tensor,
\begin{equation}
g_{\mu \nu}~\rightarrow~\Lambda_{\mu}^{\rho} 
\Lambda_{\nu}^{\sigma}g_{\rho \sigma}
+k_\mu\epsilon_\nu+k_\nu\epsilon_\mu. \label{e52}
\end{equation}
Since the 2nd and 3rd terms are vectorial, they ruin the Lorentz invariance 
and general covariance in quantum gravity, as has been specifically
pointed out by Weinberg~\cite{Weinberg72} and discussed in~\cite{Bdeo06}. 
The arguement for spin-2 case as considered in~\cite{Bdeo06} is emphasised 
again , not only as to its importance, but also to show the absence of 
spin-1 vectors which complicate calculations. Only spin-2 and spin-0 objects 
present in the theory have propagators and vertices. In string theory 
$\alpha_0^\mu=k^\mu$. So, writing
\begin{equation}
a_{\mu \nu}~|~0,k\rangle~
\rightarrow~ \Lambda_\mu^\rho\Lambda_\nu^\sigma
a^{\dag}_{\rho \sigma}~|~0,k\rangle+~O_{\mu \nu}|0,k\rangle,\label{e53}
\end{equation}
where
\begin{equation}
O_{\mu \nu}|0,k\rangle=(k_\mu \epsilon_\nu + k_\nu\epsilon_\mu)~|~0,k\rangle,\label{e54}
\end{equation}
are additional tachyons due to supersymmetry. Since $L_0=F_0^2$, we have
\begin{equation}
O_{\mu \nu}~|~0,k\rangle=L_0O_{\mu \nu}~|~0,k\rangle=
F_0^2O_{\mu \nu}~|~0,k\rangle,\label{e57}
\end{equation}
in Ramond sector. So,
\begin{equation}
F_0O_{\mu \nu}~|0,k\rangle_\alpha= \pm O_{\mu \nu}~|0,k\rangle_\alpha.\label{e58}
\end{equation}
In general, one can construct spinorial states 
$|~0\rangle_\alpha$ such that
\begin{equation}
F_0~|~0\rangle_\alpha=~|~0\rangle_\alpha~,~~~ _{\alpha}\langle 0~|~F_0
={-} _{\alpha}\langle0~|~~~\text{and}~~~ 
\sum_{\alpha}~|~0\rangle_{\alpha~~ \alpha}\langle 0~|~=1. \label{e59}
\end{equation}
Again,
\begin{eqnarray}
O_{\mu \nu}|0,k\rangle &=& L_0O_{\mu \nu}~|0,k\rangle=
\sum_\alpha F_0~|~0\rangle_{\alpha~~ \alpha}
\langle 0~|~F_0O_{\mu \nu}~|~0,k\rangle\nonumber\\
&=&-\sum|~0\rangle_{\alpha~~ \alpha}\langle0~|~O_{\mu \nu}~|~0,k\rangle=
-O_{\mu \nu}(k)=0.\label{e60}
\end{eqnarray}
This is due to the tadpole cancellation mechanism noticed also
by Casher et al ~\cite{Casher85}. Thus the Lorentz transformation
is a `proper' one and ensures that the tensor, under Lorentz transformation,
remains a symmetric tensor without vectorial components as would be 
expected from equation(\ref{e53}). Rewriting equation (\ref{g2}) 
\begin{equation}
g_{\mu\nu}(x)= \int \frac{d^3 k}{(2 \pi)^3\sqrt{(2k_0)}}
\left ( a^{\dag}_{\mu \nu}~e^{ikx} + a_{\mu \nu}~e^{-ikx}\right )\label{g2a}
\end{equation}
With operator relations (\ref{eq29}), we have quantised the tensor operator
in flat space time.
By using the same flat metric $\eta^{\mu\nu}$, we get
\begin{equation}
g^{\mu\nu}(x)= \int \frac{d^3 k}{(2 \pi)^3\sqrt{(2k_0)}}
\left ( a^{\mu \nu\dag}~e^{ikx} + a^{\mu \nu}~e^{-ikx}\right )\label{g2b}
\end{equation}
Let us consider the operator product~ :$g^{\mu\nu}(x)g_{\nu\lambda}(x)$:~.
The fermions separate out. For the two exponentials contain the factors $|z|$=1
in equation (\ref{eq7a}),
\[ \langle ~\sum_{n=1}^{\infty}\frac{1}{n}~\alpha^{-n}_{\mu}
\sum_{n=1}^{\infty}\frac{1}{n}~\alpha^{n}_{\nu}~\rangle
=\eta_{\mu\nu}~\sum_{n=1}^{\infty}\frac{1}{n}=\eta_{\mu\nu}~\zeta(1) \]
where $\zeta(s)=\sum_{n=1}^{\infty} \frac{1}{n^s}$ is the zeta function. 
This will be absorbed in the normalisation factors. It is easier to use string theory
notation so that, as in the above,
\begin{equation}
g^{ \mu\nu}(x)g_{\nu\lambda}(x)=\int \frac{d^3k}{(2\pi)^3}
\int \frac{ d^3k'}{(2\pi)^3 }~\frac{1}{\sqrt{4k_0k'_0}}~|c|^2 \left<k,0|0,k'\right>
~e^{-k.k'\zeta(1)}~f^{\mu\nu}_{~~\nu\lambda}=|c|^2
\int \frac{ d^3k}{(2\pi)^3}~e^{-k^2\zeta(1)}~f^{\mu\nu}_{~~\nu\lambda}
\left < k,0|0,k\right > =\delta^{\mu}_{~\lambda},\label{eq38}
\end{equation}
where the Kronecker delta $\delta _{\lambda}^{\mu}$ is  a mixed tensor.
The coefficient $|c|^2$ must be adjusted to have
\begin{equation}
5|c|^2\int \frac{ d^3k}{(2\pi)^3}~e^{-k^2\zeta(1)}~\left<k,0|0,k\right>=1,\label{eq38a}
\end{equation}
since
\begin{equation}
f^{\mu\nu}_{~~\nu\lambda}=\eta^{\mu}_{~\nu}\eta^{\nu}_{~\lambda}+
\eta ^{\mu}_{~\lambda}\eta^{\nu}_{~\nu}.\label{eq38b}
\end{equation}

If $g_{\mu\nu}$ is a covariant tensor, then its inverse $g^{\mu\nu}$, is 
the contravariant tensor. Other than the scalars and zero, the Kronecker tensor 
$\delta^{\mu}_{\nu}$ is the only tensor where components are same in any 
coordinate system. Furthermore, the flat space metric $\eta^{\mu\nu}$ also
satisfy $\eta^{\mu\nu}\eta_{\nu\lambda}=\delta^{\mu}_{\lambda}$. Thus we have found
a space dependant metric like the flat space one.
This meets the first criterion of Weinberg~\cite{Weinberg72} for the application of the
principle of equivalence. In the next section, we shall show that there exists a 
mapping from any curved space time point $x'$ to the point $x$ where affine is zero
so that the equations with derivatives or other equations remain the same for all space
time, flat or curved.

The relation (\ref{eq38}) is `proper' Lorentz invariant 
as per earlier extensive derivation of tadpole cancellation.
So, by the Principle of Equivalence,  $g_{\mu\nu}$ is the
true metric tensor operator for all space time points, flat or curved, 
of the gravitational field. A critical discussion on this matter has been 
given by Padmanabhan~\cite{Padma04}.

The commutator of the fields reads
\begin{equation}
\left [ g^{\mu\nu}(x), g^{\lambda\sigma}(y)
\right ]=\frac{1}{(2\pi)^3}\int\frac{d^3k}{2k_0}
\left (~e^{ik(x-y)}~-~e^{-ik(x-y)}\right )
f^{\mu\nu,\lambda\sigma}.\label{eq39}
\end{equation}
The propagator is
\begin{equation}
\Delta^{\mu\nu,\lambda\sigma}(x)=
\left<0|T\left (g^{\mu\nu}(x),~g^{\lambda\sigma}(y)\right )|0\right>=
\frac{1}{(2\pi)^4}\int d^4k~
\Delta_F^{\mu\nu,\lambda\sigma }(k)~e^{ik(x-y)},\label{eq40}
\end{equation}
where
\begin{equation}
\Delta_F^{\mu\nu,\lambda\sigma }(k)=\frac{1}{2}~f^{\mu\nu,
\lambda\sigma}~\frac{1}{k^2-i\epsilon}.
\label{eq41}
\end{equation}
We note that
\begin{eqnarray}
f^{\mu\nu,\lambda\sigma}&=&\eta^{\mu\lambda}\eta^{\nu\sigma}
+\eta^{\mu\sigma}\eta^{\nu\lambda}\nonumber\\
&=&\eta^{\mu\lambda}\eta^{\nu\sigma}
+\eta^{\mu\sigma}\eta^{\nu\lambda}-\eta^{\mu\nu}\eta^{\lambda\sigma}
+\eta^{\mu\nu}\eta^{\lambda\sigma}\nonumber\\
&=&f^{(2)\mu\nu,\lambda\sigma}+f^{(0)\mu\nu,\lambda\sigma},\label{eq40a}
\end{eqnarray}
where
\[ f^{(2)\mu\nu,\lambda\sigma}=\eta^{\mu\lambda}\eta^{\nu\sigma}
+\eta^{\mu\sigma}\eta^{\nu\lambda}-\eta^{\mu\nu}\eta^{\lambda\sigma}
~~~~\text{and}~~~~~ f^{(0)\mu\nu,\lambda\sigma}=\eta^{\mu\nu}\eta^{\lambda\sigma}. \]
The first term of equation (\ref{eq40a}) is the coefficient of the 
graviton propagator $\Delta^{graviton}_F$ and the second is the coefficient of 
dilaton propagator $\Delta^{dilaton}_F$ with appropriate factors. The Feynman
propagator for the graviton $h_{\mu\nu}(x)$ and the dilaton $D(x)$ are
\begin{equation}
\langle 0|N(h_{\mu\nu}(x)h_{\lambda\sigma}(y))|0\rangle
=\Delta^{graviton}_{F\mu\nu,\lambda\sigma}(x-y)
=\int \frac{d^4k}{(2\pi)^4}~\Delta^{graviton}_{F\mu\nu,\lambda\sigma}(k)~
e^{ik(x-y)},\label{eq40c}
\end{equation}
and
\begin{equation}
\langle 0|N(D(x)D(y))|0\rangle=\Delta^{dilaton}_{F}(x-y)
=\int \frac{d^4k}{(2\pi)^4}~\Delta^{dilaton}_{F}(k)~e^{ik(x-y)},\label{eq40d}
\end{equation}
where
\begin{eqnarray}
\Delta^{graviton}_{F\mu\nu,\lambda\sigma}(k)&=&\frac{1}{2}f^{(2)}_{\mu\nu,\lambda\sigma}
\frac{1}{(k^2-i\epsilon)},\nonumber\\
\text{and}~~~~~\Delta^{dilaton}_{F}(k)&=&\frac{1}{2}\frac{1}{(k^2-i\epsilon)}.
\label{eq40b}
\end{eqnarray}
Various  calculations, in quantum gravity, will eventually  be reduced to 
contraction of metrics $g_{\mu\nu}$'s at different space time points. The 
traceless spin-2 graviton part can be separated out with some caution. It has 
always been opined that quantisation of 
metric field strength is essential and contains even more information than the 
quantisation of gravitational field with gravitons only. For an example, let us 
construct a spin-2 graviton state, like the string state, 

\begin{eqnarray}
\Phi^{\dag}_{\mu\nu}(x)&=&
~:g_{\mu\nu}(x)-\frac{1}{2}\eta_{\mu\nu}g^{\kappa}_{~\kappa}(x):\\
&=&~:\int \frac{d^3k}{(2\pi)^3}\frac{1}{\sqrt{2k_0}}\left (a_{\mu\nu}^{\dag}-
\frac{1}{2}\eta_{\mu\nu}~a^{\dag\kappa}_{~~\kappa}\right )~e^{ikX}:\\
&\leftrightarrow &\int \frac{d^3k}{(2\pi)^3}\frac{1}{\sqrt{2k_0}}~\sum_{i,j}~c_{ij}
\left ( b_{i\mu}^{\dag}~b_{j\nu}^{\dag}- \frac{1}{2} 
\eta_{\mu\nu}b_{i\kappa}^{\dag}~b_{j}^{\dag\kappa}
\right )|0,k\rangle.\label{eq47}
\end{eqnarray}
This is a very simple way  for the string states to create the quanta of general 
relativity. One can easily check that the graviton propagator comes out correctly.
Similarly, we can write the dilaton state. The one dilaton state is
\begin{equation}
D^{\dag}(x) \leftrightarrow  \int \frac{d^3k}{(2\pi)^3}\frac{1}{\sqrt{2k_0}}
\sum_{i,j}~c_{ij}~\frac{1}{2}b_{i\kappa}^{\dag}~b_{j}^{\dag\kappa}
~|0,k\rangle.\label{eq47ac}
\end{equation}

It will be instructive, to find a projection operator which can project
out the spin two graviton from $f^{\mu\nu,\lambda\sigma}$ containing
both spin-2 graviton and spin-0 dilaton. In fact, such a projection operator
turns out to be 
\begin{equation}
{\cal{P}}^{\kappa\rho}_{~~\lambda\sigma}=\frac{1}{2}\left (
\eta^{\kappa}_{~\lambda}\eta^{\rho}_{~\sigma}+
\eta^{\kappa}_{~\sigma}\eta^{\rho}_{~\lambda}
-\eta^{\kappa\rho}\eta_{\lambda\sigma}\right ).\label{eq47ad}
\end{equation}
One can verify that
\begin{equation}
{\cal{P}}^{\kappa\rho}_{~~\lambda\sigma}f^{\mu\nu,\lambda\sigma}
=f^{(2)\kappa\rho,\mu\nu},
\end{equation}
and
\begin{equation}
{\cal{P}}^{\kappa\rho}_{~~\lambda\sigma}{\cal{P}}^{\lambda\sigma}_{\mu\nu}=
{\cal{P}}^{\kappa\rho}_{~~\mu\nu}.
\end{equation}
Further, in order to show the  graviton quanta in general relativity,let us 
apply the projection operator (\ref{eq47ad}) on the Ricci tensor $R_{\mu\nu}$
\begin{equation}
{\cal{P}}^{\mu\nu}_{~~\lambda\sigma}~R_{\mu\nu}=R_{\lambda\sigma}-\frac{1}{2}
g_{\lambda\sigma}~R=G_{\lambda\sigma},
\end{equation}
where 
\begin{equation}
R_{\mu\nu}=\partial^{\lambda}\partial_{\lambda}g_{\mu\nu}(x)+
\partial_{\nu}\partial_{\mu}g_{\lambda}^{\lambda}(x)
-\partial_{\nu}\partial^{\lambda}g_{\mu\lambda}(x)
-\partial_{\mu}\partial^{\lambda}g_{\nu\lambda}(x)
+g_{\eta\sigma}(x)\left ({\Gamma}^{\eta}_{\lambda\lambda}~
{\Gamma}^{\sigma}_{\mu\nu} - {\Gamma}^{\eta}_{\nu\lambda}~
{\Gamma}^{\sigma}_{\mu\lambda} \right ).
\end{equation}
and $G_{\lambda\sigma}$ is the Einstein's tensor of general relativity. Thus, for 
the gravitons of gravitational field of general relativity of Einstein, 
$G_{\lambda\sigma}$=0. It may be noted that we shall retain $|0,k>$ and $<k,0|$
to indicate that the vacuum of the superstring states is tachyonic.

\section{The affine connection}
There have been many successful ways of deriving Einstein's equation using the principle
of equivalence. But the study of quantum gravity implies that this principle should 
be replaceable by quantum mechanical methods which includes perturbative and 
nonperturbative expansions. The effect of full gravitational interactions can be 
obtained by replacing the ordinary derivatives by covariant derivatives or 
coderivatives. First we wish to justify the quantisation scheme with plane waves 
for the gravitational field. We shall use classical recipes.

The coderivatives of $A^{\mu}(x)$ is given by
\begin{equation}
D_{\nu} A^{\mu}(x)= A^{\mu}_{;\nu}(x)= \partial_{\nu}A^{\mu}(x)
+\Gamma_{\rho\nu}^{\mu}A^{\rho}(x).
\end{equation}
where $\Gamma_{\rho\nu}^{\mu}$ are 64 component affine connections which for 
Riemann-Christoffel tensors~\cite{Anderson67}. The transformation property, when
mapped from $x$ to $x'$, is
\begin{equation}
\Gamma_{\mu\nu}^{'\lambda}= \frac{\partial x^{'\lambda}}{\partial x^{\rho}}
\frac{\partial x^{\tau}}{\partial x^{'\mu}}\frac{\partial x^{\sigma}}{\partial x^{'\nu}}
\Gamma_{\sigma\tau}^{\rho}+ \frac{\partial x^{'\lambda}}{\partial x^{\rho}}
\frac{\partial^2x^{\rho}}{\partial x^{'\mu}\partial x^{'\nu}}.\label{eq38c}
\end{equation}
The covariant derivative of the symmetric tensor $g_{\mu\nu}$ is 
\begin{equation}
D_{\lambda}g_{\mu\nu}=g_{\mu\nu;\lambda}=\partial_{\lambda}g_{\mu\nu} -
\Gamma_{\lambda\mu}^{\rho}g_{\rho\nu}-\Gamma_{\lambda\nu}^{\rho}g_{\rho\mu}.
\end{equation}
But the affines, $\Gamma$'s and $\partial_{\lambda}g_{\mu\nu}$ vanish in local inertial
coordinate system. Once a proper tensor is zero in one coordinate system, it remains zero
in all other coordinate systems. Therefore the  covariant derivative of $g_{\mu\nu}$ is 
zero, so we get 
\begin{equation}
\partial_{\lambda}g_{\mu\nu} =
\Gamma_{\lambda\mu}^{\rho}g_{\rho\nu} +\Gamma_{\lambda\nu}^{\rho}g_{\rho\mu}.
\end{equation}
Thus all the elements of $g_{\mu\nu}(x)$ are completely determined for all points in space
time. By suitably adding three such $g$'s and using equation (\ref{eq38})
\begin{equation}
\Gamma ^{\lambda}_{\mu\nu}=\frac{1}{2} g^{\lambda\kappa}
\left (   
\partial_{\mu}g_{\kappa\nu} +\partial_{\nu}g_{\kappa\mu} -\partial_{\kappa}g_{\mu\nu}
\right).
\end{equation}
It must be remembered that the field strength has been quantised in terms of the sum of
the plane waves only. The result from such quantisation may not be valid in all 
points in the curved or flat space time. However we can argue as follows. Since both the
$g_{\mu\nu}$ and affine are symmetric, for distant parallelism for which there is well
laid down proceedure in affine geometry~\cite{Anderson67}, any of the terms of 
$\Gamma^\lambda_{\mu\nu}$ is of the form $\Gamma^\lambda_{\mu\nu}\sim - g^{\lambda\kappa}
\partial_{\mu}g_{\kappa\nu}$ and a typical derivative is $g^{\mu}_{\alpha,\sigma}\sim
-\Gamma^{\mu}_{\rho\sigma}~g^{\rho}_{\alpha}$. Further, if the affine is symmetric, 
it can be shown that ~\cite{Anderson67}
\begin{equation}
g^{\alpha\dag}_{\mu,\nu}=g^{\alpha\dag}_{\nu,\mu},
\end{equation}
and can be expressed as the gradient of four scalar $\phi_{\mu,\alpha}=
\partial_{\mu}\phi_{\alpha}(x)$, so that for another point in the space time
$x'(\mu)=\phi^{\mu}(x)$,
\begin{equation}
g^{\dag}_{\mu\nu}=\frac{\partial x'_{\mu}}{\partial x_{\nu}}, ~~~
g_{\nu\mu}=\frac{\partial x^{\mu}}{\partial x'^{\nu}}
\end{equation}
If we substitute this to calculate $\Gamma$'s of equation (\ref{eq38c}), we obtain
\begin{equation}
\Gamma^{'\lambda}_{\mu\nu}=0.
\end{equation}
Thus there always exits a mapping from $x$ to $x'$, so that affinity is flat and 
vice versa, i.e. if there is flat affinity, a particular mapping will make it a 
symmetric nonzero affinity. This justifies our use of plane wave in the Fourier 
transform for quantisation of gravitation and the result will be true for all 
space time points and for other types of expansion of the field. Most importantly 
$g_{\mu\nu}$ is the covariant and $g^{\mu\nu}$ is contravariant metric tensor of
general relativity. They are quantised in the interaction picture as given by 
equation (\ref{g2a}).

\section{ Calculation of affines, riemann and ricci tensors}
Weinberg~\cite{Weinberg72} has given a very convenient form for the
Riemann- Christoffel curvature which is useful to calculate the gravitational 
effects by replacing the ordinary derivatives. The quantum version has the 
same form
\begin{equation}
R_{\lambda\mu\nu\kappa}=~:\partial_{\kappa}\partial_{\mu}g_{\lambda\nu}(x)-
\partial_{\kappa}\partial_{\lambda}g_{\mu\nu}(x)
-\partial_{\nu}\partial_{\mu}g_{\lambda\kappa}(x)
-\partial_{\nu}\partial_{\lambda}g_{\mu\kappa}(x)
+g_{\eta\sigma}\left ({\Gamma}^{\eta}_{\nu\lambda}~
{\Gamma}^{\sigma}_{\mu\kappa}-{\Gamma}^{\eta}_{\kappa\lambda}~
{\Gamma}^{\sigma}_{\mu\nu} \right ):\label{eq98a}
\end{equation}

The Ricci tensor $R_{\lambda \mu \lambda \nu}$ is conveniently written as 
\begin{equation}
R_{\mu\nu}=R_{\mu\nu}^{(0)}+R_{\mu\nu}^{int(string)}
\end{equation}
where  
\begin{equation}
R_{\mu\nu}^{(0)}=\partial^{\lambda}\partial_{\lambda}g_{\mu\nu}(x)+
\partial_{\nu}\partial_{\mu}g_{\lambda}^{\lambda}(x)
-\partial_{\nu}\partial^{\lambda}g_{\mu\lambda}(x)
-\partial_{\mu}\partial^{\lambda}g_{\nu\lambda}(x),\label{eq99}
\end{equation}
is without gravitational interaction  and
\begin{equation}
R_{\mu\nu}^{int(string)}
=g_{\eta\sigma}\left ({\Gamma}^{\eta}_{\lambda\lambda}~
{\Gamma}^{\sigma}_{\mu\nu}-{\Gamma}^{\eta}_{\nu\lambda}~
{\Gamma}^{\sigma}_{\mu\lambda} \right )\label{eq99a}
\end{equation}
contains the interaction part.

We now proceed to find expressions for the affines. Affines as quantum operators 
should be expressions with normal ordering. The product of $g_{\mu\nu}$'s and 
their derivative contain product of two distinct pieces. All the fermion pairs 
of the affines separate out from the factors $e^{ik.X}$ as the bosonic 
$\alpha_{\mu}$'s commute with the fermionic pairs $b_{\mu,j},~b_{\nu,k}$. The
bosonic part again consists of a zero mode piece and nonzero mode piece. We shall 
use the product form which appears to be different for normal vertex forms. The zero
part of each $g_{\mu\nu}(k)$'s tachyonic part piece is
\begin{equation}
W_0(k,z)=exp{\left( k.\sum_{n=1}^{\infty}\frac{1}{n}\alpha_{-n}z^n\right )}\cdot
exp{\left(- k.\sum_{n=1}^{\infty}\frac{1}{n}\alpha_{n}z^{-n}\right )}.
\end{equation}
The correlation function is found by using
\begin{equation}
\langle \left ( \sum_{n=1}^{\infty}\frac{1}{n}\alpha_{-n}z_2^n\right )
\left ( \sum_{n=1}^{\infty}\frac{1}{n}\alpha_{n}z_1^{-n}\right )\rangle=
\eta^{\mu\nu}\sum_{n=1}^{\infty} \frac{1}{n} \left (\frac{z_2}{z_1}\right)^n.
\end{equation}
We have specified earlier that $|z_1|$=1 and $|z_2|$=1 to have the state $|0,k\rangle$,
\begin{equation}
\langle W_0(k_1)W_0(k_2)\rangle = e^{k_1\cdot k_2 \zeta(1)},
\end{equation}
Sometimes this number is left as a parameter $\lambda$. We have already adjusted this 
in equation (\ref{eq38}). Similarly
\begin{equation}
\langle W_0(k_1)W_0(k_2)W_0(k_3)\rangle = e^{(k_1.k_2+k_3.k_2+k_1.k_3) \zeta(1)},
\end{equation}
The zero mode parts are given by~\cite{Green87},
\begin{equation}
\langle Z_0(k_1)Z_0(k_2)\rangle = e^{ik_1.x}~e^{ik_2.x},
\end{equation}
and
\begin{equation}
\langle Z_0(k_1)Z_0(k_2)Z_0(k_3)\rangle=e^{ik_1.x}~e^{ik_2.x}~e^{ik_3.x}.
\end{equation}
The product $V_0=W_0~Z_0$ is then
\begin{equation}
\langle V_0(k_1)V_0(k_2)\rangle= e^{ik_1.x}~e^{ik_2.x}~ e^{(k_2.k_1)\zeta(1)},
\end{equation}
\begin{equation}
\langle V_0(k_1)V_0(k_2)V_0(k_3)\rangle= e^{ik_1.x}~e^{ik_2.x}~e^{ik_3.x}~
e^{(k_2.k_1+k_2.k_3+k_1.k_3)\zeta(1)}.
\end{equation}
and so on, as given in Ref.\cite{Green87}.
The products and derivatives of the string model metrices $g_{\mu\nu}$ satisfy simple 
commutation relations as given in equation (\ref{eq29}).
\begin{equation}
\partial^{\kappa}g^{\mu \nu}(x)~\partial_{\rho}g_{\lambda \sigma}(x)
=\int \frac{k^{\kappa} d^3~k}{{\sqrt{2k_0}}(2 \pi)^3} \frac{k'_{\rho} 
d^3~k'}{{\sqrt{2k'_0}}(2 \pi)^3}
\left ( a^{\mu \nu \dag}~e^{-ikx}-a^{\mu \nu}~e^{ikx} \right)
\left( a_{\lambda \sigma}^{\dag}~e^{-ik'x}-a_{\lambda \sigma}~e^{ik'x} \right).
\end{equation}
Because of the vacuum $|0,k\rangle$ and $\langle k',0|$ , 
\begin{eqnarray}
:\partial^{\kappa} g^{\mu \nu}(x)~\partial_{\rho} g^{\lambda \sigma}(x):~
&=&\int \frac{d^3~k}{{\sqrt{2k_0}}(2 \pi)^3} \frac{d^3~k'}{{\sqrt{2k'_0}}(2 \pi)^3}
~\langle k,0|\left[a^{\mu \nu},~a^{\lambda \sigma\dag}\right]|0,k' \rangle~ 
~e^{-k.k'\zeta(1)}~k^{\kappa}k'_{\rho}\\ \nonumber
&=&\int \frac{d^3~k}{{\sqrt{2k_0}}(2 \pi)^3} \frac{d^3~k'}{{\sqrt{2k'_0}}(2 \pi)^3}
{(2 \pi)^3}~|c|^2~f^{\mu \nu,\lambda \sigma}~\langle k,0|0,k' \rangle 
e^{-k.k'\zeta(1)} ~k^{\kappa}k'_{\rho}\\ 
&=&\int \frac{d^3~k}{(2 \pi)^3}~|c|^2~e^{-k^2\zeta(1)}~f^{\mu \nu,\lambda \sigma}~
k^{\kappa}k_{\rho}\langle k,0|0,k \rangle \\
&=&\int \frac{d^3~k}{(2 \pi)^3}\left( \eta^{\mu \lambda}\eta^{\nu \sigma}
+\eta^{\mu \sigma}\eta^{\nu \lambda} \right)~|c|^2~e^{-k^2\zeta(1)}~ k^{\kappa}k_{\rho}
\langle k,0|0,k \rangle.\label{eq99c}
\end{eqnarray}
We  use the above  equation to obtain,
\begin{equation}
\left( \partial ^{\kappa} g^{\mu \nu}(x) \right)~
\left( \partial _{\rho} g_{\lambda \sigma}(x) \right)
=\int \frac{d^3~k}{(2 \pi)^3}{k^\kappa}{k_\rho}f^{\mu \nu}_{~~\lambda \sigma}
\langle k,0|0,k \rangle.\label{eq99d}
\end{equation}                                                    

So, the general form of an affine given by
\begin{equation}
{\Gamma} ^{\lambda}_{\mu\nu}(x)=\frac{1}{2} g^{\lambda \kappa} 
\left( \partial_{\mu}g_{\kappa \nu} +\partial_{\nu}g_{\kappa \mu}
-\partial_{\kappa}g_{\mu \nu} \right),
\end{equation}
is in operator form reduces to
\begin{equation}
 :{\Gamma} ^{\lambda}_{\mu\nu}(x):~= i\int \frac{d^3 k}{(2\pi)^3}
(\eta^{\lambda}_{\nu}k_{\mu}~+~\eta^{\lambda}_{\mu}k_{\nu}~)\langle k,0|0,k \rangle
|c|^2~e^{-k^2\zeta(1)}.
\end{equation}
Here we have used the relation
\begin{eqnarray}
\int \frac{d^3 k}{(2\pi)^3}\frac{d^3 k'}{(2\pi)^3}k^{\mu}k^{'\nu}
\langle k,0|0,k \rangle\langle k',0|0,k' \rangle\nonumber
& =&\lim_{x \rightarrow 0}\int
\frac{d^3 k}{(2\pi)^3}\frac{d^3 k'}{(2\pi)^3} \partial^{\mu}\partial^{\nu}
e^{i(k-k')x}\langle k,0|0,k \rangle\langle k',0|0,k' \rangle \nonumber\\
&=& \int\frac{d^3 k}{(2\pi)^3}k^{\mu}k^{\nu}\langle k,0|0,k \rangle
|c|^2~e^{-k^2\zeta(1)} .\label{eq99b}
\end{eqnarray}
Thus we get the affines occurring in equation(\ref{eq99a}) as follows.
        
\begin{eqnarray}
 :{\Gamma}^{\eta}_{\lambda\lambda}(x):&=&i\int \frac{d^3 k}{{(2\pi)}^3}(2k^\eta)
\langle k,0|0,k \rangle |c|^2~e^{-k^2\zeta(1)},\\
 :{\Gamma}^{\sigma}_{\mu\nu}(x):&=& i\int \frac{d^3 k'}{{(2\pi)}^3}
(\eta^{\sigma}_{\nu} k'_{\mu}+\eta^{\sigma}_{\mu} k'_{\nu})\langle k',0|0,k' \rangle
|c|^2~e^{-k^{'2}\zeta(1)} ,\\
 :{\Gamma} ^{\eta}_{\nu\lambda}(x):&=&i\int \frac{d^3 k}{{(2\pi)}^3}
(\eta^{\eta}_{\nu} k_{\lambda}+\eta^{\eta}_{\lambda} k_{\nu})\langle k,0|0,k \rangle
|c|^2~e^{-k^2\zeta(1)},\\
\text{and}~~~~~~~~~~~~ :{\Gamma} ^{\sigma}_{\mu\lambda}(x):&=&i\int 
\frac{d^3 k'}{{(2\pi)}^3}(\eta^{\sigma}_{\mu} k'_{\lambda}+
\eta^{\sigma}_{\lambda} k'_{\mu})\langle k',0|0,k' \rangle |c|^2~e^{-k^{'2}\zeta(1)}.
\end{eqnarray}
Taking the product and difference of the affines at the same point `x', 
and using equation(\ref{eq99b}) we obtain  the interaction part of the Riemann 
tensor from equation(\ref{eq99a}),
\begin{equation}
R_{\mu \nu}^{int(string)}= \int \frac{d^3k}{(2\pi)^3}
\left( k^2 \eta_{\mu \nu} - k_\mu k_\nu \right)\langle k,0|0,k \rangle
 |c|^2~e^{-k^2\zeta(1)}.\label{eq99e}
\end{equation}
Let us now consider the expression, which is the remaining exact
part of the Ricci tensor
\begin{equation}
R_{\mu \nu}^{int}=~:\frac{1}{2} g_{\mu \nu} \left( \partial_{\lambda}
\partial_{\sigma} g^{\lambda \sigma}-\square g^{\gamma}_{\gamma} \right):. \label{eq99f}
\end{equation}

Using the equations(\ref{eq99c}) and (\ref{eq99d}), equation (\ref{eq99f})
can be written as
\begin{equation}
R_{\mu \nu}^{int}=\int \frac{d^3k}{(2\pi^3)}
\left( k^2 \eta _{\mu \nu}-k_{\mu}k_{\nu}\right)\langle k,0|0,k \rangle 
|c|^2~e^{-k^2\zeta(1)}. \label{eq50ac}
\end{equation}
So, 
\begin{equation}
  R_{\mu \nu}^{int(string)}=R_{\mu \nu}^{int}.  
\end{equation}
Thus the Ricci tensor of general relativity, 
\begin{equation}
R_{\mu \nu}=R_{\mu \nu}^{(0)} +R_{\mu \nu}^{int}. \label{eq50ab}
\end{equation}
is exactly reproduced in quantised form. It is perhaps one of the  rare instances 
where the first order perturbation, as suggested, gives the exact result. The 
product of operators in the expression for the affains and Ricci tensor appear 
to be independent of $x$. This is usually the case of operator product 
expansions~\cite{Polchinski98}, or the correlation function in Green function
method of calculations in quantum field theory.

\section{Conclusion}
                                                                                       
We have constructed a symmetric second rank tensor which represents the gravitational 
metric field strength. Using the equivalence of string states and field operators, 
this has been quantised in terms of the creation and annihilation operators given
in equation (\ref{g2a}). The metric quantum operator has all the properties of the 
metric of general relativity. In the interaction picture, it satisfies the weak field 
Einstein equation $R_{\mu \nu}^{(0)}=0$ for Ricci tensor, without the affines. The 
Riemann-Christoffel affines which  occur by replacing the ordinary derivatives by
coderivatives are evaluated in detail. It is found that, as another rare case, the 
first order calculation of $R_{\mu \nu}^{int}$ gives the exact $R_{\mu \nu}=R_{\mu \nu}
^{(0)}+ R_{\mu \nu}^{int}=0$ of general relativity in vacuum. The gravitational self stress 
energy in vacuum also vanishes, as shown below. The exact Einstein equation can be 
written, following Weinberg~\cite{Weinberg72},
\begin{equation}
R^{(0)}_{\mu\nu}-\frac{1}{2}\eta_{\mu\nu}R^{(0)\lambda}_{~~~~\lambda}
=-8\pi G(T_{\mu\nu}+t_{\mu\nu}),
\end{equation}
where $ t_{\mu\nu}$ is  the exact energy momentum tensor of the gravitational field itself
and is given by
\begin{equation}
t_{\mu\nu}=\frac{1}{8\pi G}\left [ R_{\mu\nu}-\frac{1}{2}g_{\mu\nu}R^{\lambda}_{~\lambda}
- R^{(0)}_{\mu\nu}+\frac{1}{2}\eta_{\mu\nu}R^{(0)\lambda}_{~~~~\lambda}\right ].
\end{equation}
This is seen to vanish due to equations (\ref{eq99}) and (\ref{eq50ab}).

So far, quantum gravity in vacuum has appeared non-renormalisable. Here we have 
attempted to show a new  way for further investigation to renormalise quantum
gravity using superstring states. Unlike most of the previous works on quantum 
gravity, we have not used a power series expansion of the metric tensor 
$g_{\mu \nu}(x)$ around a flat metric $\eta _{\mu\nu}=(-~+~+~+)$ with a coupling 
constant $\kappa$ having the dimension of length as 
\begin{equation}
g_{\mu \nu}=\eta_{\mu \nu}+\kappa \phi_{\mu \nu}.\label{eq5.1}
\end{equation}
Since the Riemann tensor and as a consequence, each term of the interaction Lagrangian
\begin{eqnarray}
{\cal L}&=& \frac{\sqrt{-g}}{16\pi G}g_{\mu\nu}^{\dag}R^{(0)}_{\mu\nu} \nonumber\\
&=&\frac{\sqrt{-g}}{16\pi G}\left [
-\partial^{\lambda}g^{\dag}_{\mu\nu}\partial_{\lambda}g_{\mu\nu}-
\partial_{\nu}g^{\dag}_{\mu\nu}\partial_{\mu}g_{\lambda}^{\lambda}
+\partial_{\nu}g^{\dag}_{\mu\nu}\partial_{\lambda}g_{\mu\lambda}
+\partial^{\mu}g^{\dag}_{\mu\nu}\partial_{\lambda}g_{\nu\lambda}
.\right ]+ \text{total derivatives}
\end{eqnarray}
contains the product of two affines, each with one  derivative and two $\phi_{\mu \nu}$'s
of equation(\ref{eq5.1}), the power counting for renormalisability comes out to 
be $N=6$. As a
result, the theory has very little chance of making quantum gravity finite by the 
usual renormalisation procedure.
When $\eta_{\mu\nu}$  is the zeroth order of $g_{\mu\nu}$ in the conventional perturbation
calculation,  one would get the graviton denoted by the Ricci tensor 
$R^{(1)}_{\mu \nu}$ as given by Weinberg~\cite{Weinberg72}. The complication 
in writing the equations for $R^{(1)}_{\mu\nu}$ 
and  $R^{(2)}_{\mu\nu}$ increases enormously; it will also be complicated , if not worse, 
in quantum gravity. Instead, by treating the metric tensor as operator function in the 
gravitational field in our case, the exact Ricci tensor is obtained in the 
first order itself.

In summary,we construct a four dimensional superstring physical state as a 
second rank Lorentz tensor equivalent operator with zero mass. These states 
are quantised as exact equivalent 
operators like the Gupta-Bleuler formalism. The metric tensor, which is not traceless, 
is found to contain both the graviton and the dilaton. The interaction 
Lagrangian, as usual, 
is the difference of the product of two affines. The product of two field strengths 
like $g^{\mu \nu}(x) g_{\nu\lambda}(x)$ which, when treated as quantum operators, 
turns out to be $\delta^{\mu}_{\lambda}$. In actual calculation, the 
two metrics at different space time are usually contracted but not the traceless 
field quanta 
of graviton alone. As a result, the  number for renormalisability of these becomes $N=4$. 
So the interaction Lagrangian theory, as formulated here, is possibly renormalisable. 
Perhaps the zeroth and first order are enough for obtaining correct results and may be a 
gift of the principle of equivalence.

It will be interesting and very much necessary to study the interaction of gravity with 
matter and radiation not only for self vacuum as we have done , and then establish 
our procedure for  renormalisation of the quantum theory of gravity with external 
energy-momentum tensor. One must also calculate graviton-graviton scattering to make 
definite statement about the renormalisability of quantum gravity interacting with matter.
It is worthwhile to point out that Mandelstam~\cite{Mandelstam75} has noted that it is only
necessary `to treat a gravitational field within itself since such a system possesses
all the essential complications of the problem of gravity'.

\end{document}